\documentclass[review]{elsarticle}
\usepackage{natbib}
\usepackage{lineno,hyperref}
\usepackage{framed,multirow}
\usepackage{amssymb}
\usepackage{latexsym}
\usepackage{makecell}

\usepackage{enumitem}

\usepackage[utf8]{inputenc}
\usepackage{amsmath}
\usepackage{amssymb}

\usepackage{color}

\usepackage{multirow}
\usepackage{colortbl}
\usepackage{hhline}

\modulolinenumbers[5]

\begin{document}

\begin{frontmatter}

\title{The Security of Deep Learning Defences for Medical Imaging}

\author[1]{Moshe Levy}
\author[1]{Guy Amit}
\author[1]{Yuval Elovici}
\author[1]{Yisroel Mirsky\corref{cor1}}

\address[1]{Department of Software and Information Systems Engineering
Ben-Gurion University of the Negev, Beer-Sheva 8410501, Israel}
\cortext[cor1]{Corresponding author: yisroel@bgu.ac.il}


\begin{abstract}
Deep learning has shown great promise in the domain of medical image analysis. Medical professionals and healthcare providers have been adopting the technology to speed up and enhance their work. These systems use deep neural networks (DNN) which are vulnerable to adversarial samples; images with imperceivable changes that can alter the model's prediction. Researchers have proposed defences which either make a DNN more robust or detect the adversarial samples before they do harm. However, none of these works consider an informed attacker which can adapt to the defence mechanism. We show that an informed attacker can evade five of the current state of the art defences while successfully fooling the victim's deep learning model, rendering these defences useless. We then suggest better alternatives for securing healthcare DNNs from such attacks: (1) harden the system's security and (2) use digital signatures.
\end{abstract}

\begin{keyword}
Medical imaging security, Deep learning, Adversarial samples
\end{keyword}

\end{frontmatter}

%\linenumbers

\section{Introduction}
\label{sec1}
\begin{figure*}[h]
    \centering
    \includegraphics[width=\textwidth]{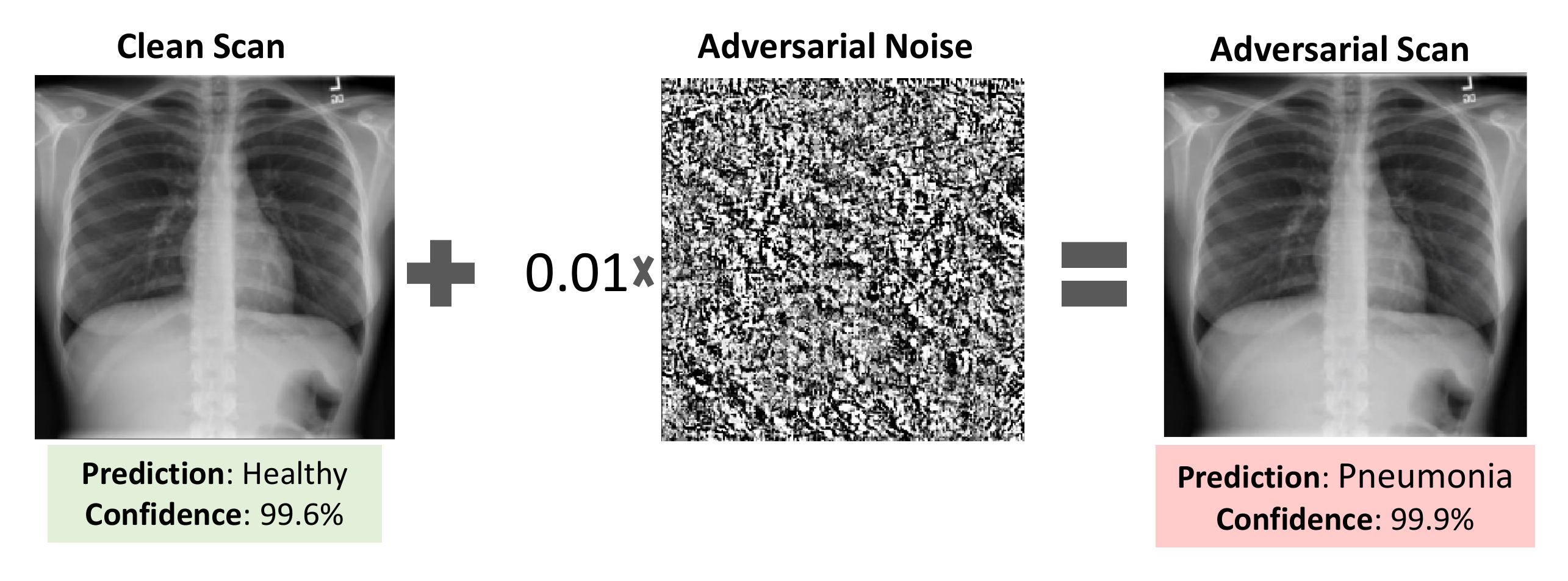}
    \caption{How adversarial noise can shift the prediction of a DNN based pneumothorax detector.}
    \label{fig:x_plus_y_is_z}
\end{figure*}
Deep learning is a data driven machine learning technique which provides state-of-the-art performance in image analysis tasks. The technique uses a model called a deep neural networks (DNN) which makes predictions by learning from historic data.
Over the past ten years, deep learning technology has proven itself as an efficient and highly accurate tool for image analysis,
solving tasks that require a wide variety of skills: from detecting cancer to spatial alignment and even content based image retrieval \citep{litjens2017survey}.
DNNs are expected to have an even more substantial role in the near future \citep{ching2018opportunities,bluemke2018radiology}.
Some deep learning solutions are already approved and deployed \citep{van2021artificial} with many more major companies in the medical field working on DNN based products, planned to be deployed soon \citep{de2020multi, bar2019improved, shadmifully}.

One of the main drawbacks of DNNs is that they are vulnerable to an attack called adversarial samples \citep{szegedy2013intriguing,7467366}. An adversarial sample is a seemingly benign input (e.g., x-ray image) which contains a crafted noise pattern, such that the pattern is imperceivable to humans but can dramatically alter the prediction of the model. For example, to change the classification of a benign x-ray to malign and vice versa (see Figure \ref{fig:x_plus_y_is_z}).

% In medical imaging, it is possible to make the DNN perform a wrong diagnosis with minimal and imperceptible adversarial noise addition.
Since DNNs are becoming mainstream in radiology and other medical imaging analysis domains, the existence of adversarial samples now threatens the healthcare community. There are number of reasons why an attacker would want to trick a medical imaging DNN. \begin{description}
\item[Earn Money.] The attacker may try to earn money. For example, quality of life insurance fraud can be achieved by inserting a lesion to a brain MRI as irrefutable evidence as to why the patient can no longer taste food. Moreover, medical scans can be held hostage in a ransomware attack where an unknown number of scans will remain tampered unless payment is made. 
\item[Cause Harm.] The goal might be to cause harm for revenge, fame, terrorism, or to cause political turmoil. For example, the attacker can cause the model to miss a lesion in an MRI or suggest the wrong diagnosis to a radiologist. 
\item[Get Priority.] An attacker may try to get medical attention faster. For example, a slight increase in the size of a bone fracture or spinal disk herniation, may lead to the patient receiving treatment sooner (over more deserving patients).
\end{description}
 
%Also, once DNN technology will have a greater foothold in medical imaging it is safe to assume more reasons will be revealed. 

Attacks on medical imagery is achievable. This is demonstrated through the million of medical records stolen in data breaches only this year\footnote{https://www.healthcareitnews.com/news/biggest-healthcare-data-breaches-2021}. Researchers have shown how Picture Archiving and Communication System (PACS) can be hacked into from the Internet \citep{McAfeeRe58:online}. Researchers have also demonstrated how easy it is to gain physical access to the PACS and plant a backdoor device in the network \citep{mirsky2019ct}. 

Since these attacks are tangible, the threat of adversarial samples has gained attention in the medical imaging community over the last few years \citep{adv_general,doi:10.1126/science.aaw4399,winter2020malicious}. As such, researchers have proposed a wide variety of detection and mitigation techniques to protect DNN-based medical imaging applications. 

\subsection{Contribution}
In this article, we warn the medical imaging community that these state-of-the-art defences provide \textbf{no security}. The reason is that all of these defences were designed assuming that the attacker will not consider the defence while crafting the attack. In reality, adversaries are adaptive and can easily evade these defences. To support our claim, we attack five state-of-the-art medical imaging defences and show that all of them fail to protect the victim's DNN from adversarial samples. We then suggest better alternatives for securing medical imaging DNNs from such attacks: (1) harden the system's security and (2) enable digital signatures for image integrity validation (a technology already supported in the DICOM standard). This article also provides an introduction to DNNs and the basic concepts of adversaries samples, at a level which approachable for individuals who are new to the domain.

By raising awareness to this issue, and by suggesting stronger countermeasures, we hope this paper will provide healthcare professionals the ability to protect their systems and patients before these attacks become more mainstream. 

%However, we identify that unique circumstances that makes the medical imaging domain especially susceptible to adversarial samples were not taken into account. 
%Mainly, that the DNN is a part of a medical AI product that the attacker can gain knowledge of, allowing for adaptive attacks that we show in this article that are substantially more effective. 
%In addition, the digital access to the imaging data allows adding the adversarial noise directly to the image file -  granting the attacker high confidence of success.

%Last paragraph
%In this article, we introduce radiologists to the basic principles of deep neural networks (DNNs) and their vulnerability to adversarial samples. 
%We discuss 5 medical imaging defenses against adversarial samples, evaluating their effectiveness under the harsh conditions of medical imaging, concluding that they are all substantially less effective than claimed. %(adaptive attacks)
%The mathematical objects and procedures are addressed in a way that is accessible to newcomers, aiming to increase awareness to the drawbacks of DNNs in medical imaging.

\label{PACS vulnerabilities}

\section{Attack Model}
In this section we describe how an attacker can gain access to medical imagery, and detail an attacker's limitations in crafting adversarial samples for these images.

\subsection{Gaining Access}

\begin{figure*}[h]
    \centering
    \includegraphics[width=\textwidth]{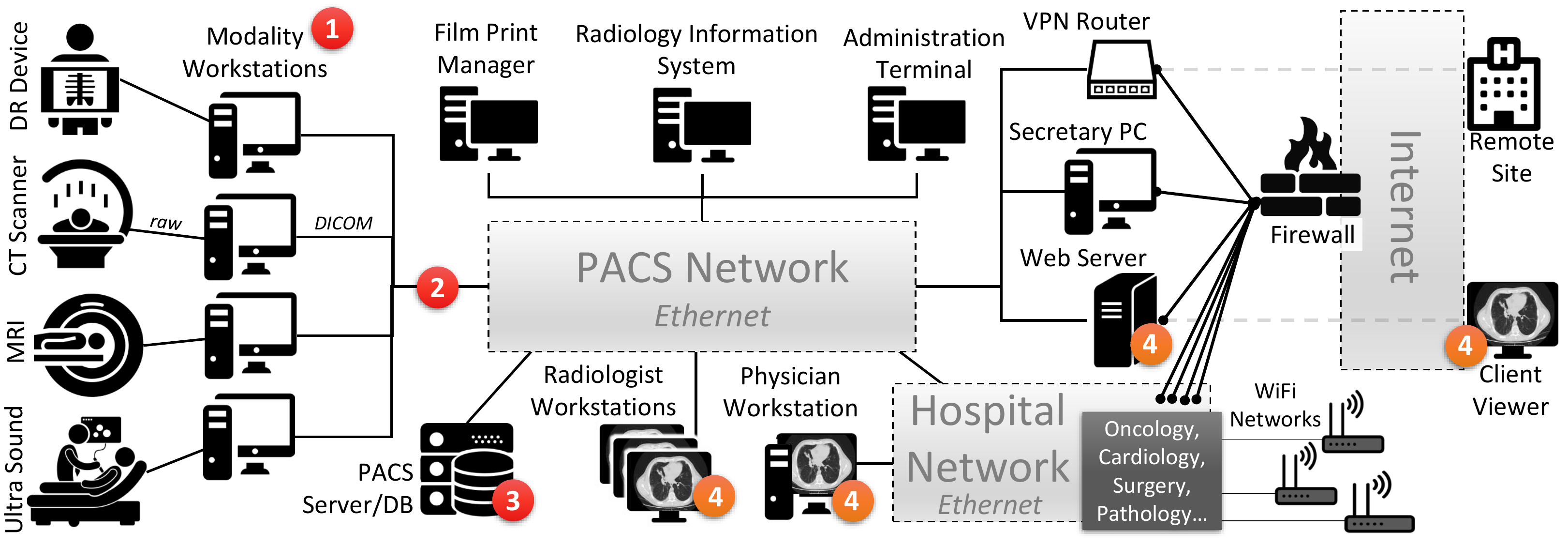}
    \caption{An overview of a hospital's PACS architecture highlighting the potential attack vectors.
    PACS is composed of 4 main components:
    (1) a secure network, (2) medical imaging devices (CT, X-ray and MRI machines), (3) Radiologist and other medical analysis workstations, and (4) a database for storing the DICOM files and reports.
    In general, there are three different attack vectors against a PACS which can get and attacker access to DICOM files \citep{mirsky2019ct}: (1) via the Internet, (2) via the institution's Wi-Fi network, and (3) via physical access to the PACS infrastructure. The labels numbered 1-5 show where the attack must take place in order to gain access to (1) images from a specific machine, (2) new scans, (3) all scans, and (4) scans associated with a specific physician or radiologist.
}
    
    \label{fig: pacs}
\end{figure*}

Medical imagery is stored in data files typically using the DICOM format. Currently, the most common way for healthcare organizations to store, manage and analyze these files is through a PACS.
The PACS provides medical personal secure access to these files from within the organization, and in some cases, from anywhere around the world.
Although the PACS network was thought to be secure, in recent years hackers have demonstrated how it can be breached both locally (on site) \citep{mirsky2019ct} and remotely (via the Internet) \citep{McAfeeRe58:online}.
In Figure~\ref{fig: pacs}, we detail the possible attack vectors against a PACS.

In 2019, over 500 healthcare organizations reported a breach impacting 23.5 million individuals. Over 2020, this figure rose over 18\% \citep{H2020rep}. That same year, one billion medical images from PACS networks were exposed \citep{Abillion19:online} with regular attacks on PACS happening in 2021 \citep{PACSVuln67:online}. Therefore, malicious access to medical imagery is open issue, one which has opened the doors to more sophisticated attacks. 
 
\subsection{Attacking the AI}\label{subsec:attackingAI}

Once the attacker has gained access to the medical imagery, he or she can convert the image into an adversarial sample by adding an imperceivable amount of noise which will fool the victim's DNN. This is in contrast to other domains where the attacker must place robust perturbations in the physical world which is significantly harder to accomplish \citep{zolfi2021translucent,Duan_2020_CVPR}.

%digital access can produce the most effective and accurate attacks. 
%Granted image file access - the attacker can prepare the adversarial noise in a safe environment, then during the attack - edit the image file itself, which is exactly the DNN's input.

However, having access to the image is not enough. To craft an adversarial sample, the attacker must have knowledge about the target DNN model. For example, explicit knowledge of the model's parameters or just abstract information on how the model was trained. The level of knowledge will impact the likelihood of a successful attack. 

\begin{description}
\item[Full Knowledge.] The attacker knows every detail about of the victim's DNN, including its trained parameters. This knowledge can guarantee the attacker a successful attack with little difficulty \citep{Carlini2017AdversarialEA,Tremer2017}. In medical imaging, the victim DNN is usually a part of an AI product which can be purchased by the attacker and then extracted, for example using the techniques of \cite{yu2020cloudleak,wei2020leaky,chen2021stealing,reith2019efficiently,tramer2016stealing,duddu2018stealing}. 

\item[Limited Knowledge.] The attacker knows how the DNN was trained (i.e., what the datasets were) but doesn't know anything about the DNN model itself. In this scenario, an attacker can train his/her own DNN model and then use this model to craft the adversarial sample. This approach is called a transfer attack \citep{Liu2017DelvingIT}. However, here, the attacker will not have a guarantee that the attack will work.

\end{description}

\section{Background}

In this section, we explain of how adversarial samples are created. We open with a brief introduction to DNNs and then describe their vulnerability.

\subsection{Deep Learning}

In general, there are several approaches for training a DNN to perform an image analysis task. The most common approach in medical imaging is the `supervised' approach where the DNN is given pairs $(x,y)$ where $x$ is an input image (e.g., a CT scan) and $y$ is the desired output of the DNN --the ground truth of $x$ for the given task. For example, for classification of lung nodules, $x$ may be an image of a lung nodule and $y$ may be a label that is either \textit{benign} or \textit{malign}. Moreover, for the task of segmentation (localization) of tumors MRI scans, $x$ may be an axial slice of the patient's head and $y$ would be a mask or probability map which indicates where the tumor is located. The collection of images and their labels used to train the DNN is referred to as the model's training set.

To train a DNN we perform the \textit{back-propagation} algorithm: (1) pass one or more examples of $x$ through the model, (2) calculate the errors between the predicted $y$ and the actual $y$, (3) propagate the errors backwards through the model to identify the erroneous parameters, (4) update (correct) the parameters using an optimization algorithm, and (5) repeat this process until model converges. The error between the actual and predicted labels is calculated using a differentiable `loss function'. The optimizer uses the gradient of the loss function to make a \textit{small step} towards the optimal solution at each iteration. This optimization algorithm is called `gradient descent`.

After training, the DNN can predict a label $y$ given an input $x$. If the training set is large and diverse, then the DNN can generalize unseen observations which were not seen in the training set.

% One of the key factors of the success of the DNN is its architecture; which specifies the order of layer and their type. Although the vast majority of implementations are use specific numerous architectures, sometimes different modifications are made to them to improve the DNN in a specific metric.

\subsection{Adversarial Samples}
\label{adversarial}
An adversarial sample $x'$ is a modified version of $x$ which includes a small imperceivable signal $\delta$, such that $x'=x+\delta$. The signal $\delta$ is crafted such that $x'$ convinces the DNN to predict the wrong label or segmentation mask for $x$ and with high confidence. The attacker who crafts $x'$ can perform a targeted attack (choose which label or mask should be predicted) \citep{goodfellow2014explaining,madry2018towards} or an untargeted attack (cause a general misclassification) \citep{kurakin2016adversarial,carlini2017towards}. What makes $x'$ dangerous is that it looks like the original $x$ to a human (see figure \ref{fig:x_plus_y_is_z}). 

The adversarial noise $\delta$ is crafted using a gradient-based technique, similar to how a DNN is trained using back-propagation. 

In general, the approach to performing an untargeted attack is to (1) pass $x$ through the trained DNN, (2) calculate the errors between the predicted $y$ and ground-truth $y$ using the model's loss function, (3) propagate the errors backwards through the model to $x$, and (4) use the loss function to update the perturbation ($\delta$) to increase the error of the victim's model. In the end, the adversarial sample is $x' = x + \delta$.
%to identify the erroneous parameters, (4) update (correct) the parameters using an optimization algorithm, and (5)(2) use the gradient of the DNN's loss function according to an image is used to craft the adversarial noise.
%The loss function can express one of two goals: in the case of untargeted attack - reducing the prediction value of the correct class \citep{goodfellow2014explaining,madry2018towards}, or in the case of targeted attack - increasing the prediction value of another class \citep{kurakin2016adversarial,carlini2017towards}. 
Researchers have shown how the attack can be improved by limiting the  magnitude of $\delta$ to make it more covert, making $\delta$ universal to different images \citep{moosavi2017universal}, and making $\delta$ robust to interference and transformations \citep{Athalye2018ObfuscatedGG}. In parallel, researchers have also proposed methods for detecting $\delta$ in $x'$ as well as decreasing its effectiveness on DNNs:
%Alongside the community effort to improve crafting adversarial samples, an extensive research was done to decrease their effectiveness. 
% A slew of defences were proposed, both as general solutions\citep{survey} that are not aimed at any domain and as domain specific solutions \citep{imagery somthing}. 
% option 2
% Since the discovery of the phenomenon in \citep{szagdy}, many works suggested defence mechanisms which decrease the DNN vulnerability to AE, making the field look like a cat and mouse game.
%--add here maybe
\begin{description}
    \item[Detection.] Defender can use an external mechanism to identify and flag potential adversarial samples. These detectors often take the form of anomaly detectors. For example, in \cite{katzir2019detecting}, adversarial samples are detected by observing abnormal behaviors within the DNN when executed on $x'$. 
    
    \item[Mitigation.] In this approach, the defender either `cleans' all input images before passing them to the DNN or changes the DNN to make it harder for the attacker to find an effective $\delta$. For example, in \cite{papernot2016distillation} the authors aimed to increase the robustness of a DNN performing a classification task. They train the DNN while forcing it to output prediction with high confidence which makes it very hard for the attacker to compute the correct gradients. 
\end{description}

It has been shown that medical images are much easier to attack than other domains \citep{ma2021understanding}. Therefore, researchers have proposed and evaluated new defence solutions for protecting DNNs used for medical image analysis. However, in 2017 researchers found that defences against adversarial attacks can be evaded by adaptive adversaries; attackers which craft $\delta$ to fool both the DNN and the defence at the same time \citep{Carlini2017AdversarialEA}. 

\section{Breaking the Defences}
\label{strategy}
In this section we describe how we collected and exploited the state-of-the-art defences in medical image analysis. To accomplish this, we took five defences for protecting medical imaging DNNs published in reputable conferences (CVPR, ISBI, MICCAI). After implementing them, we were able to break the defences by following these strategies:
\begin{description}
    \item[1. Expand the loss function.] When creating $\delta$ via back-propagation, we include both the DNN and the defence mechanism in the loss function. This approach work well with most defences which are differentiable \citep{Tremer2017}.
    \item[2. Simplify the loss function.] Sometimes, when including the defence in the loss function, it can be challenging to find the defence's mathematical derivative or to obtain a stable result. However, we can avoid these issues by substituting the defence with a simplified version and still obtain good results against the original defence \citep{Tremer2017}.
    \item[3. Calculate $\delta$ incrementally.] Although the basic approach to calculating $\delta$ involves a single pass through the DNN \citep{kurakin2016adversarial}, attack performance can be improved by (1) performing multiple passes with small update steps \citep{kurakin2016adversarial} and (2) by dynamically adjusting the step size \citep{croce2020reliable}.
    \item [4. When you fail, try again.] There are adversarial attacks that are random in nature because their initial $\delta$ is selected randomly. Therefore, when the generated $x'$ fails to fool the DNN, it is worth trying again until the best sample is created \citep{madry2018towards}.
    \item [5. Employ wisdom of the crowd.] When defence is employing a ensemble of DNNs or when there is limited knowledge of the victim DNN it can be useful to use a set of DNNs to craft the attack \citep{he2017adversarial}.
\end{description}

We now detail each of the five medical imaging DNN defence methods and how we successfully crafted adversarial samples which evade their protection. 

\subsection{MGM Method}
\textbf{Defence Method.}
In~\cite{li2020robust} the authors propose the MGM detector. 
MGM is a detector that analyzes the internal behavior of the DNN when processing a sample. It assumes that adversarial samples will induce an abnormal behavior in the outputs of the final hidden layer. By fitting these outputs to a Gaussian distribution on a population of clean samples, they are able to identifying any deviations (adversarial samples).

\noindent\textbf{Exploitation.} 
To perform a successful attack, we must bypass the detector while fooling the DNN. To achieve this, we generate adversarial samples using both the victim DNN and MGM in our loss function. However, the Gaussian likelihood function of MGM makes the crafting of adversarial samples numerically unstable. Instead, we found that if we craft adversarial samples using a simplified version of the MGM (with an approximated Gaussian likelihood function) the adversarial samples can fool the original MGM as well:

\noindent Original MGM function (multivariate Gaussian):
\begin{equation}\label{eq1}
\begin{split}
     \mathcal{L} & = \log\big(\mathcal{N}(x; \mu,\Sigma)  \big) \\
    & = -\frac{d}{2}\log(2\pi) -\frac{1}{2}\log(|\Sigma|) -\frac{1}{2}\big\Vert x-\mu \big\Vert^{2}_{{\Sigma}^{-1}}    
\end{split}
\end{equation}
Our approximated MGM function:
\begin{equation}\label{eq2}
     \mathcal{\hat L} = \big\Vert x-\mu \big\Vert^{2}_2
\end{equation}

%Main contribution here 
As seen above, our main finding is that one can simply take the L2 distance from the center of the learned Gaussian distribution to effectively and efficiently craft adversarial samples that bypass the MGM detector.

\subsection{GMM Method}
\textbf{Defence Method.} 
In \cite{li2021defending}, the authors perform both prevention and detection in their solution. For prevention, the DNN is trained on both normal and adversarial samples to make it more robust to attacks, an idea that has been shown to be effective in the past \citep{szegedy2013intriguing}. 
%The intuition behind this combination is that the prevention mechanism makes it harder to craft adversarial samples and forces the attacker to create samples which can be more easily identified by the detector.
%It does so by adding a set of adversarial samples to the DNN training data, this is believed to encourage the DNN to make more robust predictions that are less affected from adversarial noise.
For detection, the approach is similar to that of MGM; the detector observes the outputs of the last hidden layer and assumes a Gaussian distribution. However, in contrast to MGM, GMM detector uses a separate Gaussian distribution for each prediction class of the classifier.
%This detection approach was evaluated on the OCT dataset~\citep{kermany2018identifying}, and used a performance measure called "adversarial risk"(detail in~\ref{sec:results}).
%After the victim DNN makes its prediction, the Gaussian likelihood corresponding to the predicted class is used for adversarial detection. The likelihood of the Gaussian is used as a measure of confidence that the sample is clean and is associated with the predicted class. 
%If the likelihood exceeds a threshold, the sample is deemed clean, otherwise it is tagged as adversarial.

% the detector evaluates its confidence that the sample is clean, if the confidence is crossing a set threshold, the sample is tagged as clean. The confidence is the likelihood of the Gaussian that is corrospoding to the class that was predicted by the victim DNN.

\noindent\textbf{Exploitation.} 
Because of the prevention mechanism, we were not able to bypass the defence like we did for MGM. Instead, we evaded the prevention-detection system by using a two stage adversarial crafting process: 
In the first stage we find the ``nearest'' class which fools the victim model, and the second stage we convince both the victim DNN and detector that it is the correct class and untampered. To accomplish this, we first lower DNN's confidence of $x'$ having the true label $y$. Then, after several training epochs, we update the perturbation to increase both the detector and DNN's confidence that $x'$ the input is safe belongs to the attacker's target class.
% predicting that , the attack checks to which of the classes the sample is associated with from the detector perspective, and

\subsection{Ensemble Method}
\label{sec: ensemble}
\textbf{Defence Method.} 
To challenge the attacker, the authors of~\cite{paul2020mitigating} use ``wisdom of the crowd`` by using multiple victim DNNs and by taking their average prediction. Doing so increases the complexity for the attacker.
% Using an ensemble usually improve performance and makes the predictions more robust to small changes in the input.
To further enhance the overall robustness of the ensemble, the authors include adversarial samples in their training set.

In this paper the victim DNNs are trained on a private dataset containing 2D slices of CT images.
The slices were selected by a radiologist such that they will capture best the nodule in the image.
To reproduce this work, we used the CT images from the from the LUNA16 dataset~\citep{setio2017validation}, which also include radiologist annotations of slices similar to the ones used in the paper.

\noindent\textbf{Exploitation.} 
Although the ensemble increases the complexity of generating an adversarial sample,  we found that we don't need to attack all of the DNNs in parallel. Instead, we found that an attacker can fool all of the DNNs by (1) generating an adversarial perturbation for each DNN separately, (2) averaging all of the perturbations together, and (3) applying the averaged perturbations to the image. 
We found that this attack also works using different perturbation methods. We used  a one step attack (FGSM~\citep{goodfellow2014explaining}) since it was able to fool the entire ensemble while being the simplest.
Formally, we performed the following attack:
\begin{equation}\label{eq3}
\begin{split}
    &\hat \nabla_{avg} = \frac{1}{N}\sum_{i=1}^{N}sign(\nabla_{x}\mathcal{L}_{i}(x, y)) \\
    &x_{adv} = x + \epsilon \cdot \hat \nabla_{avg}
\end{split}    
\end{equation}
Where $\nabla_{x}\mathcal{L}_{i}(x, y)$ is the $i^{th}$ DNN's loss gradient with respect to the input $x$, where the label is $y$, $\epsilon$ is scaling hyper parameter, and $sign()$ is the sign function which maps positive values to 1 and negative values to -1.
In other words, to overcome the``wisdom of the crowd'', we only needed to exploit each model individually. 
%Autoencoder
%

\subsection{Denoiser Method}
\begin{figure*}[h]
    \centering
    \includegraphics[width=0.8\textwidth]{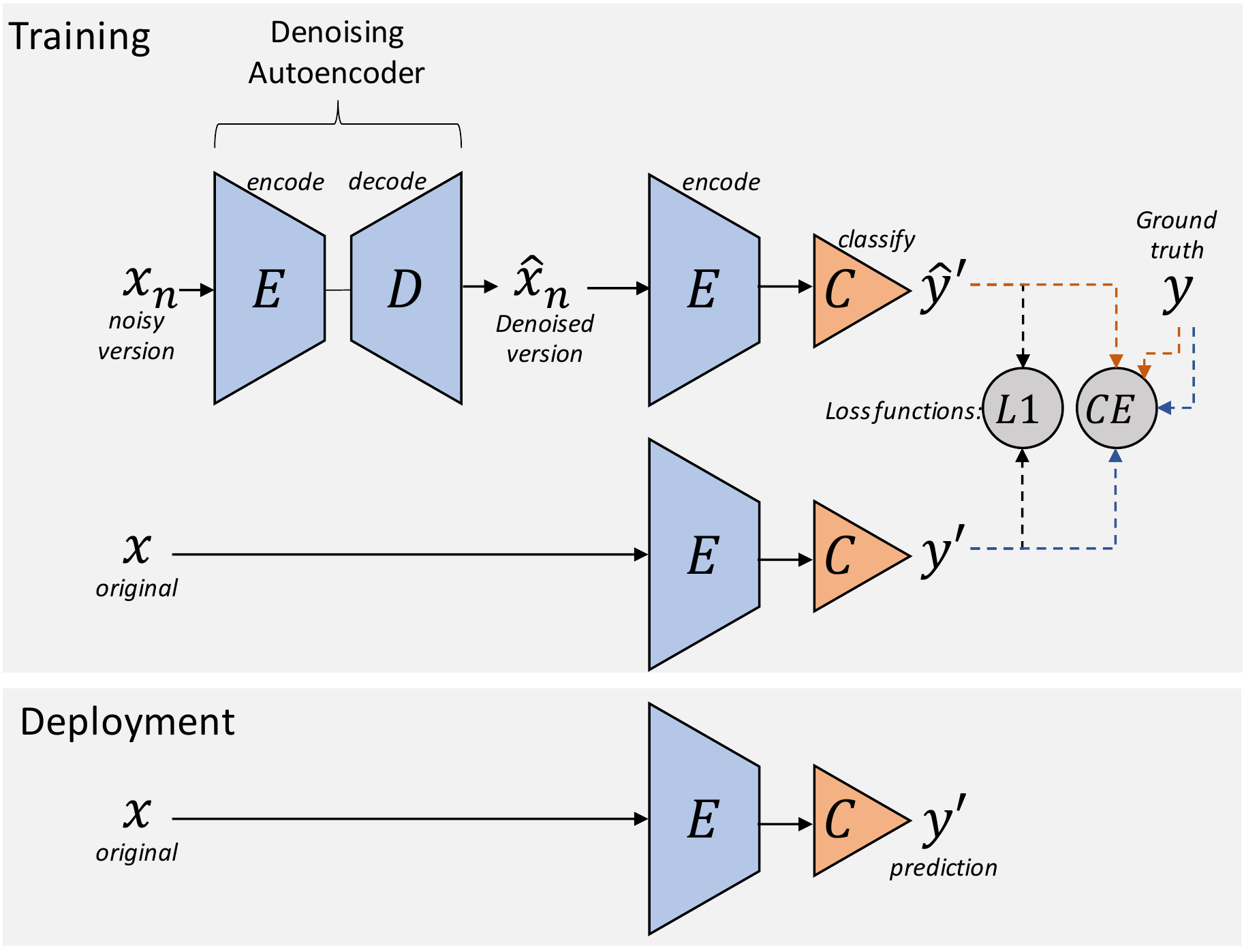}
    \caption{The schematic for the Embedded Denoiser defence. During training, noised version $x'$ of the input $x$ is passed through a denoising autoencoder ($E$-$D$). Both $x$ and $x'$ are then passed through a classifier which utilizes the denoising-encoder's representation of $x$. The entire model is trained to (1) minimize the discrepancy between noisy are clean images using L1 loss, and (2) minimize the error in classification using cross entropy loss (CE).
    }
    \label{fig:AE}
\end{figure*}

\textbf{Defence Method.}
In \cite{xue2019improving}, the authors suggest embedding a denoising autoencoder neural network into the victim's DNN to mitigate noise such as adversarial perturbations. A diagram of the architecture is provided in Figure~\ref{fig:AE}.
The DNN is trained to both optimize the classification and  increase the model's resilience to noise(such as adversarial noise).
In the paper, the authors do not claim adversarial robustness in a white box scenario, instead they perform evaluations for limited knowledge attacks.
%The defence mechanism is evaluated on two datasets: one for skin lesion detection (SKIN4) and one for x-ray pneumonia detection(RSNA Pneumonia).
\\
\noindent\textbf{Exploitation.} 
Because this paper claims to defend against limited knowledge scenarios, we attack the model using the surrogate DNN approach (see section \ref{adversarial}). 
First we created a small ensemble of DNNs which all use the defence architecture. Then we used the attack depicted in~\ref{sec: ensemble} to generate adversarial samples.
Finally, we select the most effective adversarial samples and evaluate them on the victim's actual DNN.

\subsection{RBF Method}
\textbf{Defence Method.} 
To mitigate attacks, the authors of \cite{taghanaki2019kernelized} increase a DNN's robustness by modifying the victim DNN's architecture. This is accomplished by adding a layer after each block of convolutions. The layer applies a radial basis function (RBF) to the concatenation of the block's input and the convolutional layer's output \citep{taghanaki2019kernelized}. This approach follows the finding that RBFs have greater robustness than standard DNNs found in \cite{goodfellow2014explaining}.
%The authors claim this RBF layer hardens a DNN against adversarial samples. 
%The authors of the paper suggested this defense included evaluations on skin lesion detection (SKIN4) and one for x-ray diagnosis(CHEST14).

% Alt version: 
% This prevention method suggests to modify the DNN architecture in a way that should increase its robustness. It adds a layer after certain layers, claiming this specific layer diminish issues in the DNN behavior that cause it to be vulnerable to adversarial samples. The layer is added on top of existing layers, meaning that its output is concatenated to the output of the layer that is placed on top of. The layer's function is detailed in \citep{CVPR}
\noindent\textbf{Exploitation.} 
The RBF layer indeed increases the difficulty of crafting adversarial samples. However, we were able to evade the RBF layers through patience and diversity: (1) samples were generated using many iterations with a slow learning rate, and (2) for each $x$ we used different attack algorithms and selected the best one as suggested in \cite{croce2020reliable}.

\section{Defence Evaluation}\label{sec:eval}
In this section we empirically measure the security of the defences by attacking them with the exploits described in the previous section. First we provide our experiment setup and then present the impact which our attacks have on the defences and the respective victim models.

\subsection{Datasets}
\begin{enumerate}
    \item \textbf{CHEST14:} a
    public X-ray dataset gathered by the NIH Clinical Center.
    This dataset contains more than 100,000 anonymized X-ray scans from 30,000 patients. 
    The scans are accompanied by annotations for 14 types of medical conditions.
   In our paper we use two versions of the dataset. CHEST14 and CHEST2, where CHEST2 only indicates whether a scan has some medical condition.
   
   \item \textbf{RSNA-X-ray:} a public dataset published on Kaggle~\footnote{https://www.kaggle.com/c/rsna-pneumonia-detection-challenge.} by the Radiological Society of North America (RSNA).
   This dataset contains 30,000 chest X-rays of patients with and without Pneumonia.
    
   \item \textbf{CT-Slices:}  CT images from the from the LUNA16~\citep{setio2017validation} lung cancer dataset, which includes annotations from radiologists.
   We used the annotations to crop benign and malign nodules from the scans.
   
   \item \textbf{Brain MRI segmentation dataset:} a collection of FLAIR MRI scans from the Cancer Imaging archive (TCIA). The scans consist of 110 patients included taken from the Cancer Genome Atlas (TCGA) lower-grade glioma collection.
   These scans are accompanied with FLAIR abnormality segmentation masks and genomic cluster data.
   The dataset can be found on Kaggle~\footnote{https://www.kaggle.com/andrewmvd/brain-tumor-segmentation-in-mri-brats-2015}.
   
  \item \textbf{OCT data:} a public retinal optical coherence tomography (OCT) database, which was introduced in~\cite{kermany2018identifying}.
  The database contains 84,495 images from 4686 patients, divided into 4 classes according to their medical condition.
  For our evaluation, we followed the image sampling process described in~\cite{li2021defending}.
  
  \item \textbf{ISIC}: the skin lesion dataset from IEEE ISBI International Skin Imaging Collaboration (ISIC) Challenge described in ~\citep{codella2018skin}.
  The dataset includes two types of annotations: lesion segmentation masks and disease categories,
  we use both of these annotations in our evaluation.
   
\end{enumerate}

% In Table~\ref{table:datasets} we present a summary of the datasets used for the evaluation of each defence method.

% \input{tables/datasets}

\subsection{Metrics}
We measure the impact of an attack on a defence by measuring the drop in the defence's performance. We used a different performance metric for each machine learning task: 

\begin{description}
    \item[Classification.] For binary classifiers we use accuracy and for multi-class classifiers we use average-accuracy. We compute GMM's accuracy using the inverse of the author's metric called `adversarial risk' --a combined performance measure of the detector and the victim's classifier \citep{li2021defending}. Additionally, for MGM\cite{li2020robust} and GMM\cite{li2021defending}, which used an evaluation different from accuracy, we performed the exact same evaluation made in the original papers. In MGM, we use the AUROC metric that measures the area under the ROC curve and in GMM we used the adversarial risk (directly).
    
    \item[Segmentation.] We used the Dice measure which expresses how much the ground-truth segmentation map fits (overlaps) the predicted one. The metric has values on the range of 0-1 where higher is better.
\end{description}

\subsection{Experiment Setup}
All of our code was written using the Pytorch 1.7.0 framework. Both the models and attacks were performed using Nvidia 3090 GPUs.
\begin{description}

\item{\textbf{Defences Implementation.}}\label{implementation} For each of the defence methods, we used the exact same architecture as proposed in the respective paper except for two cases: (Ensemble Method) The ensemble classifier proposed in \cite{paul2020mitigating} performed poorly on our available medical data (without attacks). Therefore, we used the popular Resnet20 architecture which is similar in size and performed better \citep{he2016deep}. (RBF Method) The RBF layer method from \citep{taghanaki2019kernelized} achieved low performance on our clean datasets and we could not obtain authors implementation. Therefore, we used the popular Resnet34 which is similar in size \citep{he2016deep} and improved the victim DNN's performance. For all of the defences, we measured their performance on clean data (not attacked) as a baseline for the attack performance.  
\item{\textbf{Attacks Experiments.}} For each defence method, we first reproduced the authors' results by attacking the defence with the original adversarial attack. We obtained a similar defence performance as reported by the authors, indicating that our implementations are correct. We then attacked the defences using our new attacks. We bounded the adversarial noise energy to the same level used by the authors (making the attacks invisible to humans). All boundaries were set by the norm $l_\infty$.
\end{description}

\subsection{Results}

\begin{table*}[t]

\centering
\caption{The performance of five state of the art radiological DNN defence methods against our adaptive attacks. A higher is value means a safer DNN. We compute the accuracy of GMM by taking $100$ -  \textit{adversarial risk}. The MGM detector is calibrated to detect 5\% of the clean samples as adversarial. The `Original Attacks' are the adversarial attacks used in the respective papers. In summary, an adaptive adversary can craft adversarial samples which can evade all of the defences and fool the victim model.\\}

\resizebox{.8\textwidth}{!}{
\begin{tabular}{cc|c|c|c|cc|}
\multicolumn{1}{l}{}         & \multicolumn{1}{c}{} & \multicolumn{1}{l|}{} & Baseline               &  \begin{tabular}[c]{@{}c@{}}Attacked w/o \\Defence\end{tabular}              & \multicolumn{2}{c|}{Attacked with Defence}                                                                                                            \\ 
\cline{3-7}
\multicolumn{1}{l}{}         &                      & Dataset               & \textit{Clean Samples} & 
\begin{tabular}[c]{@{}c@{}}\textit{Original}\\ \textit{Attack}\end{tabular} & \begin{tabular}[c]{@{}c@{}}\textit{Original}\\ \textit{Attack}\end{tabular} & \begin{tabular}[c]{@{}c@{}}\textit{Our}\\ \textit{Attack}\end{tabular}  \\ 
\hline
\multirow{10}{*}{\rotcell{}} &                      &                       & \multicolumn{4}{c|}{{\cellcolor[rgb]{0.937,0.937,0.937}}Accuracy (Classification)$\uparrow$}                                                                                                                                                                 \\
                             & MGM                  & CHEST14               & 95.2                   & 0                                                                           & 90.4                                                                        & 4                                                                       \\
                             & Ensemble             & CT-Slices             & 84.7                   & 0                                                                           & 80.4                                                                        & 0                                                                       \\
                             & Denoiser             & RSNA-X-ray            & 85.5                   & 30.1                                                                        & 43.2                                                                        & 3                                                                       \\
                             & RBF                  & CHEST2                & 84.6                   & 62.3                                                                        & 50                                                                          & 22.9                                                                    \\
                             & GMM                  & RSNA-X-ray            & 79.5                   & 0                                                                           & 45.4                                                                        & 19.4                                                                    \\
                             & GMM                  & OCT                   & 99                     & 0                                                                           & 52.1                                                                        & 14.1                                                                    \\ 
\hhline{~------|}
                             &                      &                       & \multicolumn{4}{c|}{{\cellcolor[rgb]{0.937,0.937,0.937}}Dice (Segmentation) $\uparrow$}                                                                                                                                                                      \\
                             & RBF                  & Brain MRI             & 82.9                   & 23.9                                                                        & 39.8                                                                        & 18.5                                                                    \\
                             & RBF                  & ISIC                  & 77.6                   & 43.9                                                                        & 83.3                                                                        & 22.1                                                                    \\
\hline
\end{tabular}}
\label{table:results}
\end{table*}

\begin{table*}
\centering
\caption{The performance of MGM and GMM against adaptive attacks measured using the same metrics from the original articles. The Table confirms that we have successfully reproduced the works and that the attacks harm these metrics as well. Ensemble, Denoise and RBF are omitted from this table because Table \ref{table:results} already presents their original metrics.\\}
\label{table:results_sup}
\resizebox{.8\textwidth}{!}{
\begin{tabular}{cc|c|c|c|cc|}
\multicolumn{1}{l}{}        & \multicolumn{1}{c}{} & \multicolumn{1}{l|}{} & Baseline               & \begin{tabular}[c]{@{}c@{}}Attacked w/o \\Defence\end{tabular}              & \multicolumn{2}{c|}{Attacked with Defence}                                                                                                            \\ 
\cline{3-7}
\multicolumn{1}{l}{}        &                      & Dataset               & \textit{Clean Samples} & \begin{tabular}[c]{@{}c@{}}\textit{Original}\\ \textit{Attack}\end{tabular} & \begin{tabular}[c]{@{}c@{}}\textit{Original}\\ \textit{Attack}\end{tabular} & \begin{tabular}[c]{@{}c@{}}\textit{Our}\\ \textit{Attack}\end{tabular}  \\ 
\hline
\multirow{5}{*}{\rotcell{}} &                      &                       & \multicolumn{4}{c|}{{\cellcolor[rgb]{0.937,0.937,0.937}}AUROC (Classification)$\uparrow$}                                                                                                                                                                    \\
                            & MGM                  & CHEST14               & 87.1                   & 12.5                                                                        & 96.9                                                                        & 0                                                                       \\ 
\hhline{~------|}
                            &                      &                       & \multicolumn{4}{c|}{{\cellcolor[rgb]{0.937,0.937,0.937}}Adversarial risk (Classification)$\downarrow$}                                                                                                                                                       \\
                            & GMM                  & RSNA-X-ray            & 20.5                   & 100                                                                         & 54.6                                                                        & 80.6                                                                    \\
                            & GMM                  & OCT                   & 1                      & 100                                                                         & 47.9                                                                        & 85.9                                                                    \\
\hline
\end{tabular}}
\end{table*}
\label{sec:results}
In Table~\ref{table:results} we provide a summary of our experiment results. The results show that the models perform as expected on clean samples and reproduce similar results when attacked with a non-adaptive adversary (as reported by the authors). However, as an adaptive adversary, for each model we were able to craft adversarial samples which both evaded the respective defence and fooled the victim's classifier (or segmentation model). Overall, our attacks reduce a defended victim's accuracy by 50\% on average for classification and 30\% for segmentation. In three cases, the attacks reduce the models' performance to 0-4\% (i.e., fooling the models for nearly every possible input $x$).

The three most vulnerable defences are the MGM, Ensemble, and Denoiser methods which are completely broken in the case of adaptive adversaries. These methods were the most vulnerable because we were able compute gradients over their models. In contrast, the RBF and GMM methods were slightly more robust to our adaptive attacks because their gradients were harder compute. However, they were still defeated overall with performance dropping by 30\% on average (rendering the system untrust worthy). We also note that both the Ensemble methods and GMM use adversarial training to make their models more robust (where the model is trained using adversarial samples as well).
However, this approach does not provide a guarantee of security \citep{madry2018towards}, as evident from the fact that Ensemble method was completely defeated. We note that on the X-ray dataset, the RBF defence seems to make it \textit{easier} for the attacker to craft samples (evident from the decrease of accuracy). This may have to do with the way the authors connected the RBF layer for the task of classification. We reached out to the authors several times for comment and have not received any response. In summary, all of the defence methods are vulnerable to gradient attacks which enable adaptive attackers to bypass the defences. 

In order to verify that our results reflect the authors' original implementations, we examine our results using their metrics. The results in the original metrics for RBF, Ensemble and Denoiser can be found in Table~\ref{table:results} and MGM and GMM can be found in Table~\ref{table:results_sup}. By comparing the baseline results to those reported in the original papers, we conclude that (1) our implementations of their papers are correct and (2) that their defences are indeed vulnerable.

There is a general principle that security by obscurity does not provide protection. Based on our results and analysis, we conclude that these state-of-the-art defences do not provide protection for the same reason. Attackers with full-knowledge and even limited-knowledge can exploit vulnerabilities in these defences to evade detection and prevention.

\section{Discussion}

The fundamental issue with the current defences is that they do not prevent an attacker from observing their gradients signals. This is known to be an open problem in the AI community \citep{Tremer2017,shamir2021dimpled}. Another way of looking at it is that the machine learning community, including researchers in medical imaging, are stuck `fighting fire with fire', trying to solve machine learning problems with machine learning tools. As a result, these defences suffer from the same oversight (they assume a static attacker). To prevent the users of these technologies from being misled, we recommend that researchers attack their own defences and include the analysis in their work \citep{Tremer2017}. 

In the presence of an adaptive adversary, defenders of DNN models are indeed at a disadvantage. However, unlike other domains, the medical imaging community has two advantages which it can use to guarantee the security of its DNNs:
\begin{description}
    \item[Closed Environment]  In general, there are many opportunities for an attacker to tamper a sample before it reaches a DNN. %This is especially true in other domains, such as malware detection where the attacker has full control over the initial sample. 
    This is especially since the attacker usually has full control over the initial sample (e.g., malware detection, autonomous driving, and face recognition). 
    However, unlike other domains, medical imaging networks (PACS) are closed systems. This means that most of the attack vectors listed in section~\ref{PACS vulnerabilities} can be mitigated by hardening the system's security: employing standard security measures (e.g., network traffic encryption) and by performing regular system updates. As a result, the attacker will not be able to access samples to perform the attack. Therefore, we suggest that PACS administrators should focus more on securing their network and end-devices than employing adversarial attack detection models.
    
    \item[End-to-end Attribution] %Hardening the security of a PACS network will mitigate the attacks but not prevent them. This is because absolute cybersecurity is hard to achieve (e.g., due to human error and zero-day attacks). 
    Another advantage medical imaging has over other domains is that the PACS has access to all medical media over its entire life-cycle (from creation to analysis). As such, it is possible to deploy a technology called digital signatures which can provably guarantee that pixels or metadata of an image (i.e., a DICOM file) have not been tampered. Digital signatures work as follows: an entity (e.g., the CT scanner) (1) takes a hash of the document producing a summary code $m$, and then (2) encrypts $m$ using a private one-way encryption key, producing the signature $s$. Then, all other entities (radiologist workstations or DNN analysis tools) can verify that the document has not been tampered by (1) decrypting $s$ using a public one-way decryption key, and then (2) checking that the result is identical to the hash of the document. The reason this process guarantees the integrity of the signed document is because (1): if the document were tampered, the signature/hash would not match, and (2) nobody else can encrypt messages that work with the public key aside from the scanner (who has the private key). The DICOM standard has a field which can hold digital signatures, and some modality scanners support their creation \citep{Dicomdig}. Therefore, we suggest that all PACS networks enable digital signatures where possible, and more importantly, verify them at the end-points before performing any analysis.

\end{description}

In conclusion, we found that all state-of-the-art defences for DNNs in medical imaging do not provide any security since an adversary can craft attacks which both evade the defence and fool the DNN model. Our analysis revealed that the core issue of these defences is that they do not prevent adversaries from exploiting the victim's gradient. Since this is an open issue, the defenders appear to be at a disadvantage. However, medical imaging networks are at an advantage over other domains: they can mitigate access to medical scans, and they can employ digital signatures to guarantee image integrity. 

We hope this research will (1) help medical professionals and healthcare providers understand the threat of adversarial samples in image analysis, (2) help medical researchers avoid the pitfalls of developing ineffective defences, and (3) help PACS administrators make informed decisions on how to secure their networks from these attacks in the future.

\section*{Acknowledgements}
\noindent This material is based upon work supported by the Zuckerman STEM Leadership Program.
This project has received funding from the European union's Horizon 2020 research and innovation programme under grant agreement 952172.

\begin{figure}[h]
    
    \includegraphics[width=.1\textwidth]{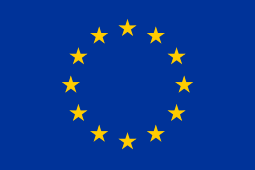}
    %\captionsetup{labelformat=empty}

\end{figure}

\bibliography{mybibfile}

\end{document}